\documentclass[letterpaper,aps,prl,twocolumn,showpacs,preprintnumbers]{revtex4}
\usepackage{graphicx}
\newcommand{\mybm}[1]{\mbox{\boldmath$#1$}}

\begin{document}

\title{Diluted Magnetic Semiconductor at Finite Temperature}
\author{Shih-Jye Sun}
\affiliation{Department of Applied Physics, National Chia-Yi University, Chia-Yi 600, Taiwan}
\author{Hsiu-Hau Lin}
\affiliation{Department of Physics, National Tsing-Hua University, Hsinchu 300, Taiwan}
\affiliation{Physics Division, National Center for Theoretical Sciences, Hsinchu 300, Taiwan}
\date{\today}
\begin{abstract}
We studied the diluted magnetic semiconductor by the self-consistent Green's function approach, which treats the spin-wave kinematics appropriately at finite temperatures.  Our approach leads to a simple formula for the critical temperature in a wide range of parameter space. In addition, the magnetization curve versus temperature in some regimes is concave, which is dramatically different from the usual convex shape. Finally, we discuss the possibility of generalizing the current theory to include the realistic band structure, electronic correlations and disorders in a systematic way.
\end{abstract}
\maketitle

Diluted magnetic semiconductor (DMS) has attracted intense attentions\cite{Ohno98a,Prinz98} for its potential applications in spintronics devices. Robust ferromagnetic order has been observed in (Ga$_{1-x}$Mn$_x$)As up to 110 K\cite{Ohno96,Ohno99}. Magnetically doped wide bandgap semiconductors and oxides such as GaN, ZnO, TiO$_2$ even exhibit ferromagnetism at room temperature\cite{Matsumoto01,Theodoropoulou01,Lee02}, although the magnetization is less robust than the doped III-V semiconductors. While lots of efforts are focused on the search of optimal materials with enhanced critical temperatures\cite{Datta90}, it remains a challenging task to describe the coupled localized moments and the itinerant carriers in an appropriate way. 

Part of the difficulty lies in the fact that the spatial fluctuations of the ferromagnetic order is large at finite temperature and the usual Weiss mean-field theory does not work. To account for the spatial fluctuations, numerical approaches\cite{Dietl00,Schliemann01,Sham01,Sun02}, such as local density functional approximation and Monte Carlo simulations, are quite helpful in estimation of various thermodynamical properties. However, it is rather difficult to study the electronic transport by these numerical approaches. Analytical approaches\cite{Akai98,Konig00,Litvinov01,Berciu01} provide partial descriptions in several particular limits by treating the impurity spin semiclassically or ignoring the interactions and the kinematic constraints among spin waves. At finite temperatures, the average spin-wave density is large so that these approximations are no longer appropriate. It is therefore desirable to develop a spin-wave theory which works at finite temperature.

In this Letter, we adapt the self-consistent Green's function approach to describe the fluctuating spin correlations at finite temperature. Since the kinematic constraint of spin waves are treated exactly in the equation of motion, this method can be applied to a wide range of parameter space, even very close to the critical temperature\cite{Vogt01,Yang01}. Indeed, the critical temperature is determined by the simple formula,
\begin{equation}
k_{B} T_c = \frac{S+1}{3} \left[\frac{1}{cV} \sum_{p} \frac{\alpha_{c}}{\Omega_{c}(p)}\right]^{-1},
\label{CriticalT}
\end{equation}
where $c$ is the density of impurity spins. The spin-wave dispersion $\Omega_{c}(p)$ and the impurity spin polarization $\alpha_c$ are both determined self-consistently near the critical temperature $T \to T_c$. The trends of the critical temperature upon the change of parameters are studied in detail later.

We model the DMS by the Hamiltonian, containing only the kinetic energy of itinerant carriers and the exchange interaction between the itinerant and the localized impurity spins,
\begin{equation}
H =H_0 + J \int d^{3} r\: \mybm{S}(r) \cdot \mybm{s}(r), 
\end{equation}
where $J > 0$ is the strength of exchange interaction. The impurity spin density is $\mybm{S}(r)=\sum_{I} \delta^{(3)}(r-R_I) \mybm{S}_{I}$ while the itinerant spin density is $\mybm{s}(r) = \psi^{\dag}(r) (\mybm{\sigma}/2) \psi(r)$. The band structure of the itinerant carriers is described by $H_0$, which depends on the host semiconductors. Since our emphasis here is how to cope with spatially fluctuations appropriately, the dispersion is taken as the simplest parabolic band, $H_0 = p^2/2m^{*}$. Generalization to more realistic but complex band structures, such as the Luttinger model, can be achieved straightforwardly.

Since the impurity spins are randomly doped into the host semiconductor, their positions are random. If the disorder is strong, the itinerant electrons are localized and the percolation approach\cite{Litvinov01,Berciu01} would be more appropriate. However, we are interested in the metallic regime where itinerant carriers are delocalized. The disorder also plays a crucial role in smoothing out the impurity spin density $\mybm{S}(r)$. For instance, the magnitude of the spin density is smeared after coarse-graining,
\begin{equation}
\langle \mybm{S}^2(r) \rangle_{R_I} \approx c^2 S(S+1),
\end{equation}
where $c$ is the impurity spin density and the average is taken over random locations of the impurity spins. Therefore, the presence of weak disorder allows a field-theory description in the continuous limit.

The dynamics of the impurity spins is described by the thermal Green's function,
\begin{eqnarray}
D(r,\tau) &\equiv& \langle\langle S^{+}(r,\tau); S^{-}(0,0) \rangle\rangle 
\nonumber\\
&\equiv& -\Theta(\tau) \langle [S^{+}(r,\tau), S^{-}(0,0)] \rangle. 
\end{eqnarray}
The equation of motion for $D(r, \tau)$ would involve more Green's functions of higher orders. The exact solution then involves the Green's functions of all orders and is not feasible in general. However, within mean-field approximation, the higher-order Green's functions can be decomposed into simpler ones and eventually a self-consistent solution is possible. For the spin-wave propagator $D(q,i\nu_n)$ in momentum space, the mean-field decomposition simplifies the equation of motion,
\begin{equation}
i\nu_n D(q, i\nu_n) = 1+ J \langle s_z \rangle +\frac{J\langle S_z \rangle}{c^* V} \sum_{k} F(k,k+q,i\nu_n),
\label{EOM1}
\end{equation}  
where $c^*$ is the density of itinerant carriers.
Notice that it only involves one additional Green's function $F(k,k+q,i\nu_n) \equiv \langle\langle \psi^{\dag}_{\uparrow}(k) \psi_{\downarrow}(k+q); S^{-}(0,0)\rangle\rangle$. Applying the same trick again, the equation of motion for $F(k, k+q, i\nu_n)$ is
\begin{equation}
F(k, k+q, i\nu_n) = \frac{Jc^*}{2} \frac{f_{\uparrow}(k) -f_{\downarrow}(k+p)}{i\nu_n +\Delta + \epsilon_{k} - \epsilon_{k+p}} D(q, i\nu_n),
\label{EOM2}
\end{equation}
where $f_{\uparrow, \downarrow}(k) = [e^{\beta(\epsilon_{k} \pm \Delta/2 -\mu)}+1]^{-1}$ is the Fermi distribution for itinerant carriers with different spins. The Zeeman gap $\Delta \equiv J \langle S_z \rangle$ in the electronic band structure is due to the ferromagnetic order of the impurity spin and has to be determined self-consistently.

\begin{figure}
\centering
\includegraphics[width=8cm]{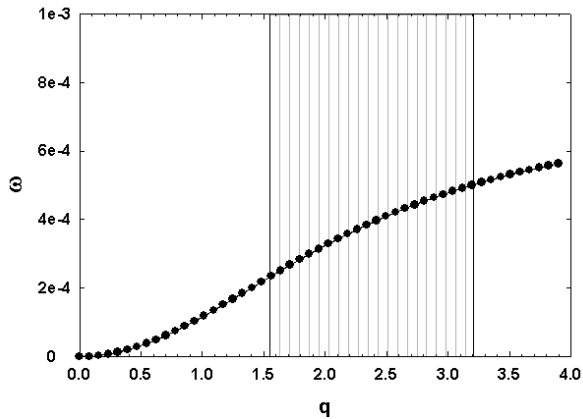}
\caption{\label{f1} The magnon excitation spectrum at zero temperature. The
impurity density $c=1.0/{nm^3}$ and $c^*=0.01/{nm^3}$. The shaded area represents the regime where the imaginary part of the self energy is not zero but negligibly small.}
\end{figure}

From Eqs.~(\ref{EOM1}) and (\ref{EOM2}), we can solve for the spin-wave propagator. Upon the Wick rotation, $i \nu_n \to \Omega + i\eta$, the spin-wave dispersion is identified as the simple pole in the Greens' function,
\begin{equation}
\Omega - J \langle s_z \rangle -\frac{J\Delta}{2V} \sum_{k} \frac{f_{\uparrow}(\epsilon_{k}) -f_{\downarrow}(\epsilon_{k+p})}{\Omega +\Delta + \epsilon_{k} - \epsilon_{k+p} + i\eta} =0.
\label{Dispersion}
\end{equation}
The average spin densities $\langle s_z \rangle$ and $\langle S_z \rangle$ in the above equation remain unknown and need to be determined self-consistently.

The average itinerant spin density $\langle s_z \rangle$ is just the difference of the spin densities in majority and minority bands, split by the Zeeman gap $\Delta$. The relation between the average localized spin density $\langle S_z \rangle$ and the spin-wave dispersion $\Omega(k)$ is more subtle due to the non-trivial spin kinematic constraint. From Callen's formula,\cite{Callen63}
\begin{equation}
\frac{1}{c}\langle S_z \rangle = S - \langle n_{sw} \rangle + \frac{(2S+1) \langle n_{sw} \rangle^{2S+1}}{(1+ \langle n_{sw} \rangle)^{2S+1} - \langle n_{sw}\rangle^{2S+1}}.
\label{Magnetization}
\end{equation}
Here $\langle n_{sw} \rangle = (1/cV) \sum_{k}[e^{\beta \Omega(k)}-1]^{-1}$ is the average number of spin waves over all momenta. At low temperatures, the average spin-wave number is small and the last term in Eq. (\ref{Magnetization}) can be safely ignored. The magnetization show the $T^{3/2}$ behavior as in the independent spin-wave approximation. As the temperature approaches the critical regime, the spin-wave density becomes large and the kinematic constraint becomes important. Solving Eqs. (\ref{Dispersion}) and (\ref{Magnetization}), both the spin-wave dispersion $\Omega(k)$ and the averaged impurity spin density $\langle S_z \rangle$ are obtained self-consistently. 

The dispersion from Eq. (\ref{Dispersion}) has two branches due to the presence of both itinerant and localized spins. Because the gapless spin-wave fluctuations dominates, the optical branch can be safely ignored. Besides, the spectral weight of the optical mode is also small due to the dilute density of itinerant carriers. In Fig. \ref{f1}, the lower branch of spin-wave dispersion at zero temperature is shown, corresponding to the Goldstone excitation of the ferromagnetic order. Our numerical results show that the imaginary part of the spin-wave self energy, due to the presence of St\"oner continuum (the optical branch), is negligible in all regimes and the dispersion is sharp. Thus, it is a reasonable approximation to assume the spectral weight is carried by the gapless spin-wave excitations only.

\begin{figure}
\centering
\includegraphics*[width=8cm]{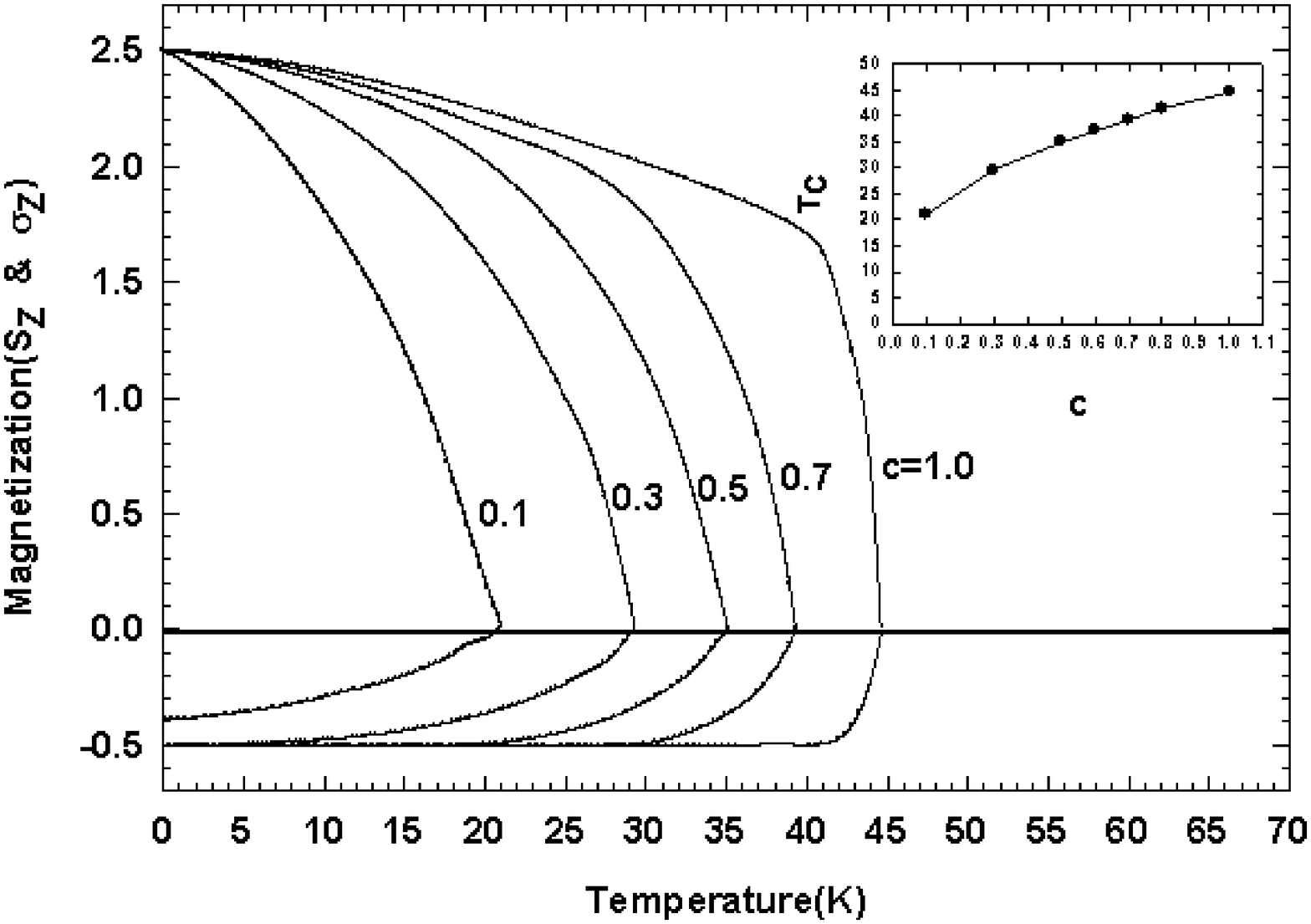}
\caption{\label{f2} Temperature dependence of spin polarizations at different
impurity densities $c$. The ratio of itinerant and impurity spin densities are fixed at $c^{*}/c =0.1$ for all curves. The inserted figure shows the monotonically increasing critical temperature $T_c$ at different densities.}
\includegraphics*[width=8cm]{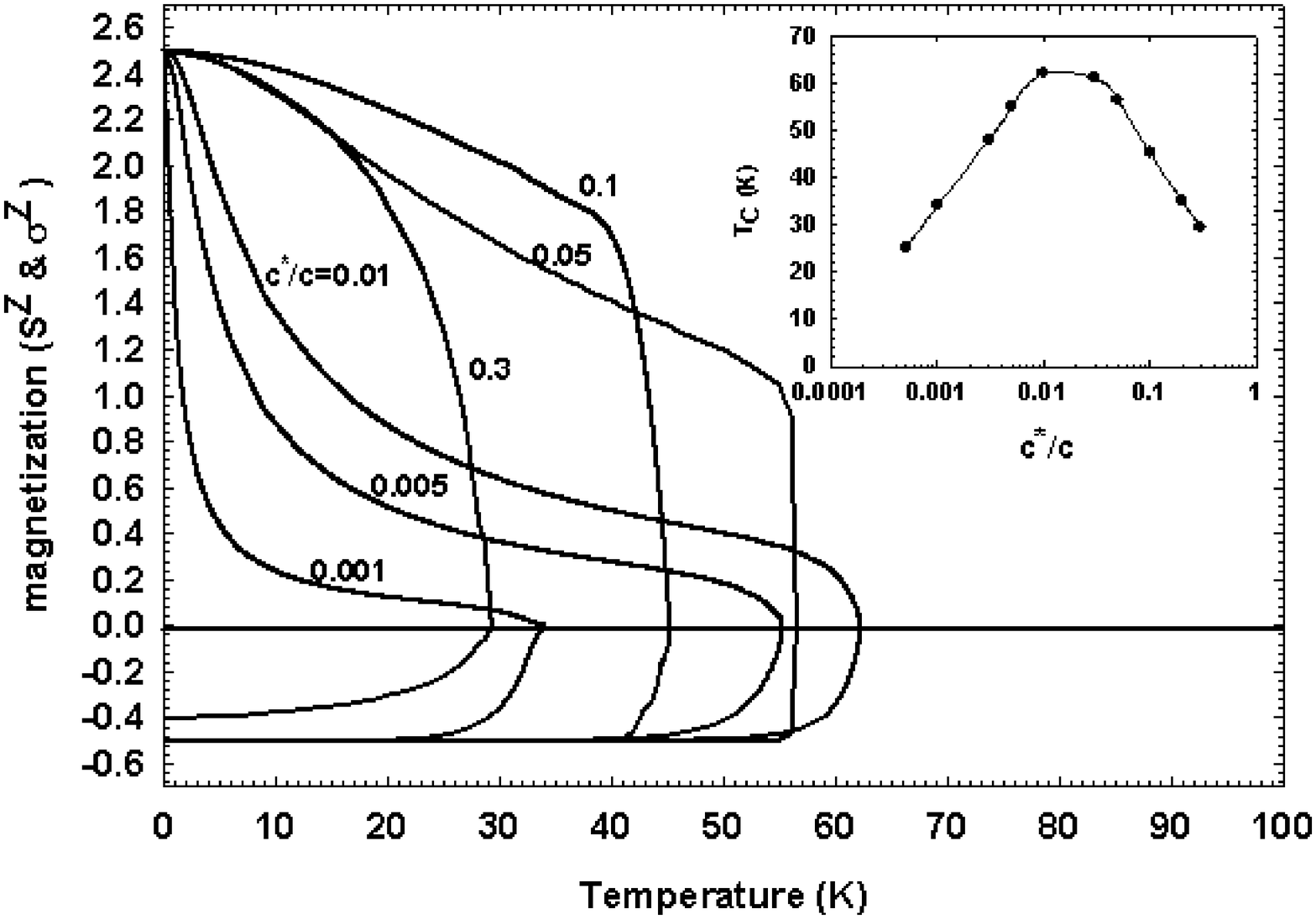}
\caption{\label{f3} Magnetization curves at different itinerant spin densities $c^*$, while the impurity spin density is fixed at $c=1$ nm$^{-3}$. The critical temperature $T_c$, as shown in the inset, reaches the maximum at an optimal concentration of the itinerant spin density.}
\end{figure}

The magnetization curves at different spin densities $c^{*}, c$ are plotted in Figs. \ref{f2} and \ref{f3}. Throughout this Letter, the exchange coupling and the effective mass are fixed at typical values $J = 0.15$ eV nm$^3$ and $m^* = 0.5 m_e$\cite{Ohno99}. First of all, we study the temperature dependence of the magnetization at different impurity spin densities $c$, while the ratio of itinerant and localized spin densities fixed at $c^*/c =0.1$. It is quite interesting to notice that the magnetization drops more dramatically near the critical regime when the impurity spin density is large. As shown in Fig.~\ref{f2}, the critical temperature, computed by Eq.~\ref{CriticalT}, increases monotonically as the densities increase\cite{Sun03}. This monotonic increase in the critical temperature is qualitatively (not quantitatively) the same as the Weiss mean-field theory and also agree with the experiments. 

On the other hand, if the impurity spin density is hold constant, say $c=1$ nm$^{-3}$, both the magnetization curve and the trend of the Curie temperatures show interesting behaviors, which deviate from the Weiss mean-field theory.
Starting from extremely dilute density of the itinerant carriers, the magnetization curve is concave as shown in Fig.~\ref{f3}, in contrast to the usual convex curvature in Weiss mean-field theory. This concave shape resembles the magnetization curve in percolation theory near the boundary of the metal-insulator transition and is often used to be an indication of localization in the presence of disorder\cite{Litvinov01,Berciu01}. However, our results show that a concave magnetization curve is not necessarily tied up to the localization tendency. We emphasize that the disorder of impurity spin is included in our approach only through the coarse-graining procedure and the transport of the itinerant carriers is assumed ballistic here. So it is rather surprising, by including the spin-wave fluctuations appropriately at finite temperatures, the magnetization curve is concave at dilute densities.

In the regime where the magnetization curve is concave, the critical temperature increases with the carrier concentration $c^*$. Beyond an optimal density, $c^* \sim 0.01$ nm$^{-3}$ (the impurity spin density is fixed at $c = 1$ nm$^{-3}$), the critical temperature reaches the maximum and starts to fall back as in the inset of Fig.~\ref{f3}. The suppression of the critical temperature is mainly due to the oscillatory RKKY interaction at higher concentrations. The effective coupling between impurity spins is frustrated and not longer purely ferromagnetic. The existence of an optimal density is totally missed in the Weiss mean-field theory, where both the spatially varying fluctuations and the quantum frustrations are ignored.

Beyond the optimal density, not only the critical temperature falls back, the shape of the magnetization also undergoes an interesting change. There are two significant features. One is the magnetization curve becomes convex again. Another feature is that the magnetization has a very sharp decrease near the critical temperature, making it look almost like the first-order phase transition. To make sure that the sharp drop is not an artifact of numerical errors, the data points near this regime are chosen very closely to ensure we still have enough points in the steep regime. In particular, for $c^* =0.05$, it is spectacular that the magnetization drops 50\% within 1K. It is not clear at this point what is the underlying mechanism behind this dramatic suppression of magnetization. A more sophisticated theory starting from the critical point might be able to address this interesting dive of magnetization near the critical temperature. 

\begin{figure}
\centering
\includegraphics*[width=8cm]{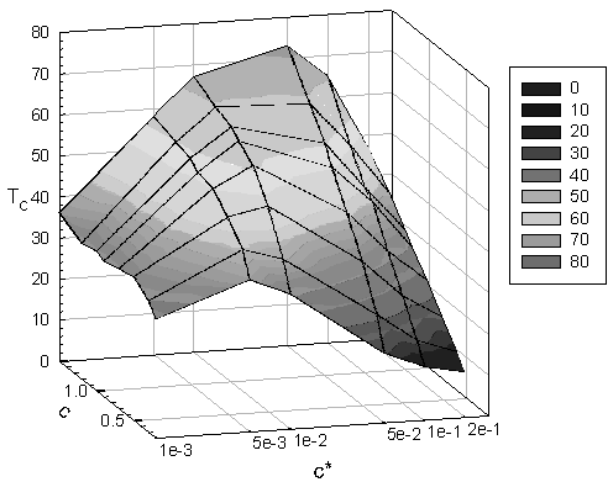}
\caption{\label{f4} The trend of the critical temperatures at different densities.}
\end{figure}

Since the global trend of the critical temperature is important, it is plotted in Fig.~\ref{f4}. Comparing with Fig. 2 in Ref.\cite{Dietl00}, the self-consistent Green's function approach produces a more complicated landscape. In the Weiss mean-filed theory, the critical temperature is
\begin{equation}
T^{\rm MF}_c = \frac{\chi_{P}}{(g^{*} \mu_B/2)^2} \frac{S(S+1) J^2 N}{12}, 
\end{equation}
where $g^*$ is the $g$-factor of the carriers and $\chi_{P}$ is their Pauli susceptibility, which is proportional to the effective band mass. The above formula would produce a profile with monotonically increasing $T_c$ as the densities become larger. It is clear from Fig.~\ref{f4} that the profile of the Curie temperature has a ridge, roughly along the curve $c \sim 100 c^{*}$ with the given parameters in this Letter, and is {\em qualitatively} different from previous studies. A closer check would find that our approach also produces quantitative differences. Comparing with the estimates of  Curie temperatures, previously studied by one of the authors in Refs. \cite{Schliemann01} and \cite{Konig00}, the present approach gives a lower $T_c$ in all densities regimes. So it seems that the inclusion of spin kinematics at finite temperatures is not only important to get the right trend of the critical temperatures, but also crucial in estimating their values.

Finally, we address the important aspects of physics which are left out so far. To make quantitative comparison with the experiments, it is crucial to adapt the realistic band structure of the host semiconductors, for instance, the six-band Luttinger model for GaAs. The inclusion of the more complex band structure would not cause formidable messes in calculating the Green's function self-consistently. While  most  of the conclusions drawn from the simple parabolic band should remain valid, a more realistic band structure is desirable for making quantitative predictions. The electronic correlation can also be included at the mean-field level in a systematic way. Since Coulomb repulsive interaction stabilizes the ferromagnetic phase, the Curie temperature is expected to be higher. To include the disorder is more subtle. In the diffusive regime, one can replace the ballistic propagator of the itinerant carriers by the diffusive one to account for the disorder effects.

In conclusion, we employ the self-consistent Green's function approach to study the DMS and derive a general formula for the Curie temperature. In addition,  we demonstrate the interesting crossover of the magnetization curve from concave to convex, which is not driven by the localization effects.

We thanks Chung-Yu Mou and Ming-Fong Yang for fruitful discussions and the support of National Science Council in Taiwan. The hospitality of the National Center for Theoretical Science, where the work was initiated, is greatly acknowledged.

\end{document}